# Energy efficient routing in mobile ad-hoc networks for Healthcare Environments

**Sohail Abid[1], Imran Shafi[2] and Shahid Abid[3]**

**[1] Department of Computing and Technology**
**IQRA University, Islamabad, Pakistan**

**[2] Department of Computing and Technology**
**Abasyn University Islamabad campus, Pakistan.**

**[3] Foundation University Institute of Engineering**
**and Management Sciences, Pakistan**

**Abstract**

The modern and innovative medical applications based on wireless network are being developed in the commercial sectors as well as in research. The emerging wireless networks are rapidly becoming a fundamental part of medical solutions due to increasing accessibility for healthcare professionals/patients reducing healthcare costs. Discovering the routes among hosts that are energy efficient without compromise on smooth communication is desirable. This work investigates energy efficiency of some selected proactive and reactive routing protocols in wireless network for healthcare environments. After simulation and analysis we found that DSR is best energy efficient routing protocol among DSR, DSDV and AODV, because DSR has maximum remaining energy.

***Keywords:*** *Energy Efficient Routing in healthcare environment, Energy Awareness, energy-efficiency, medical application, Simulation of Energy Efficient Routing Protocol in NS2.*

## 1. Introduction

Today Mobile Ad-hoc Network (MANET) is a rapidly growing technology, due to its unique nature of distributed resources and dynamic topology. The routing protocols in MANET have standards which controls the number of nodes that harmonies the way to route packets between all the mobile nodes in their networks. There are lots of challenges to MANET like security, network stability, performance, efficiency and energy efficiency etc. Now a day wireless MANET's are becoming very popular and many routing protocols have been suggested by researchers. We give brief introduction on applications of wireless networks in the medical field and discuss the issues and challenges regarding to performance and efficient use of energy. We are concerning with energy efficiency and select some well known energy efficient routing protocols and simulate these protocols in NS2 and analyze energy efficiency in different cases.

Most of the healthcare equipments which are equipped with Wireless technology need energy efficient routing. It is vital in healthcare environments because The Energy efficient routing protocols were introduced years ago. Today Energy efficient routing protocols are used in wireless MANET and WSN. A Wireless Mobile Ad-hoc Network (MANET) is a set of mobile nodes that are randomly and dynamically placed in such a way that the interconnections between hosts are proficiently changing on a regular basis. The energy efficient routing protocol is used to discover routes between hosts to smooth the progress of communication within the network and try to utilize minimum energy consumption. The primary objective of such a MANET energy efficient routing protocol is best, correct and efficient route establishment between a pair of nodes so that messages may be delivered in a timely manner and save energy. A minimum overhead and bandwidth consumption should be done in the creation of route and maintenance of route [1].

Now a day mobile ad hoc networks have focused much more attention to the convenience of building mobile wireless networks without any need for a pre-existing infrastructure. A mobile ad hoc network is a group of wireless mobile nodes which are capable and agree to establish relations, using without any centralize supervision and infrastructure [2]. Mobile ad-hoc networks provide an environment, in which each node acts as a router for example receives packets and forwards to the nearest node or next hop, in order to reach final destination through various hops.

Wireless technology is ideal for medical applications and equipment due to its everywhere accessibility and mobility. Most of the heavy and expensive machines have limited mobility. These machines are not being fully used for a long time. Wireless technologies provide new interfaces to these heavy machines and make them interrelate with any new machines and use it everywhere. Some wireless technologies which are used in medical field are CodeBlue, MobiHealth and project connect etc. [3], [4]. Different medical applications for hospital have been developed like hospital management, equipment management and patient management etc. due to use of these applications efficiency of hospitals is increasing. Wireless technology for healthcare applications and equipment is rapidly growing. Some of these wireless medical equipments which are being used in hospitals are ECG anywhere, Life Source Products, Health Trax, LifeSync Wireless ECG System etc. [4].

The wireless networking has a bright future in the field of medical applications. Now day access and cost reductions are two hottest issues in the field of medical or healthcare. Both of these ends are successfully achieved by wireless networking. Around the world medical care organizations are rapidly getting complex, especially in the United States. Nearly 98000 patients die every year due to preventable medical errors. Wireless Network provides tools that can help reduce such medical errors. In wireless applications which are used in medical field, one of the major issue is availability of power. It is being guaranteed that the routing protocol is energy efficient and show full performance in less energy environment.

### 1.1 Mobile Ad-Hoc Network Challenges

Mobile ad-hoc network uses a broad range of applications. In real world for the impact of these applications, we require professional, reliable and more efficient algorithms and protocols. The MANET faces lot of challenges, which understand carefully and unmistakably [5]. Some of these challenges are discussed below:

Quality of Service: It is also important factor that the packet of data which is sent by source reached at destination timely and reliably. The protocol quality of service (QoS) is very critical in some applications like audio, video streaming.

Scalability: In ad-hoc network it is most important when a network is extended or expanded the protocol handle it smoothly and reliably. The protocol is flexible enough to respond and operate with such large number of hosts.

Ad-hoc Deployment: In a particular area ad-hoc network deployment is different according to the application. In ad-hoc network hosts are randomly deployed in the region without any knowledge of topology and prior infrastructure. In this situation the distribution of nodes and identification of connectivity between nodes are depend upon nodes.

Fault-Tolerance: In unfriendly environment, a host may fail due to certain problems or lack of power or energy. If a host fails, it is the responsibility of the protocols to accommodate these changes in the network.

Physical Resource Constraints: Limited battery power is most important and challenging constraint forced on MANET network host. The power supply is determined by MANET host directly. The energy consumption is the main issue in MANET.

In this paper, we focus on the MANET energy-efficient routing techniques regarding the network protocols that have been developed in recent years.

## 2. Background

### 2.1Ad-Hoc Routing Protocol

Three different types of ad-hoc routing protocols describe in Fig. 1, details are as under.

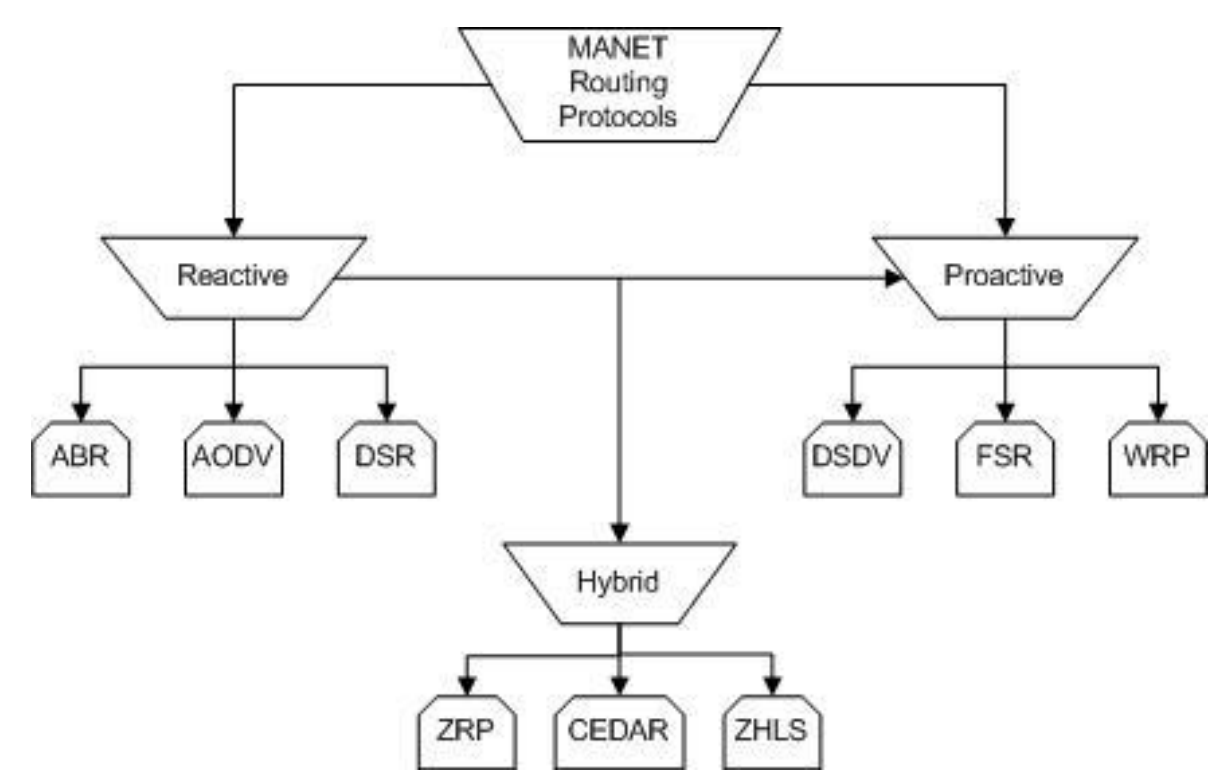

Fig. 1: Mobile ad-hoc routing protocols

#### 7. Proactive Routing Protocol

In proactive routing protocols, first routing table is created and then data has to be sent from one host to another. These protocols regularly sends request to their neighbor nodes to organize their network topology and as well as make the routing table.

#### 7. Reactive Routing Protocol

In this type of protocols when a host sends data to another host, it asks their neighbor nodes for a route. If neighbor nodes have no route, they broadcast the request to their neighbor nodes and so on.

### 7. Hybrid Routing Protocol

These protocols are the combination of reactive and proactive protocols. The important achievement in hybrid protocol is to minimize latency and broadcast. In hybrid protocols the network is divided into many small parts and each part has a gateway, inner part use reactive routing and inter-part use proactive routing.

In this paper DSDV is simulated and compared with reactive routing protocols AODV and DSR. The main goal of this task is to examine how to find different routing techniques impact the energy consumption in MANET. The summary of the three routing protocols are given below.

### AODV

The process of rout discovery in AODV (Ad-hoc On-demand Distant Vector Routing) is that it initiate route discovery and initially the routing information of the table is empty. First of all a host or a node send / broadcast RREQ (route request) or acknowledgement (Ack) packet to all its neighbor nodes. The RREQ packet is a collection of broadcast ID, source address, source sequence no., destination address, destination sequence no. and total no. of hop count. AODV algorithm is appropriate for a dynamic user-starting network like when user wants to utilize ad-hoc network. AODV gives free of loop routing environment still when recovering broken links.

### DSDV

DSDV (Destination-sequenced Distance Vector) was developed by C. Paerkins in 1994. Bellman-Ford algorithm is used in DSDV. This algorithm was developed for graph search applications and this is also used for routing purpose. Like every table-driven protocol, when the data has to be sent from source to destination, the DSDV minimize the latency by having a route. In the network each host maintain a routing table, the routing table count hops and tell how many hops arrived from source to destination. The routing table in DSDV protocol consists of Seq. no, hop count and route. The updated routing table is sent to each and every host of network in order to update table regularly.

### DSR

DSR (Dynamic Source Routing) is working on the concept of source routing and on-demand routing protocol [6]. Each host is mandatory to maintain route caches which restrain the source routes that are acknowledged to it. The route caches are constantly updated by the host as it continues to find out new routes.

It has two major operations, first route discovery and second route maintenance. If a packet is sent to a destination node, the node will check with its own route cache for its obtainable route to the destination which is not expired. If a route present, the packet will be sent through the existing route. But, if a route does not present, it will send RReq (Route Request) to all host in the network. This is a point where route discovery phase start. The Route Request (RREQ) packet contain of unique identification number, source and destination address. When route request reached to a host, the host compare the new route to its own routing table, if does not match, it will add in routing table and forward the packet to it neighbor. This process will continue until all nodes in the network have a route to the destination.

## 3. RELATED WORK

Different routing protocols have been produced by the researchers with the help of simulation software. Some of them have also been used to minimize the energy consumption. L. M. Feeney presented in his paper a comparison of energy consumption for DSR, AODV in NS2 [7]. The analysis considers the cost for sending and receiving traffic, for dropped packets, and for routing overhead packets. Fr´ed´eric Giroire and his team present a link which connects the two routers. The two network interfaces join via this link [8]. Their goal is to find new routes that reduce links between source and destination while completing all requirements. Li Layuan, Li Chunlin, and Yuan Peiyan presents energy level based routing protocol “ELBRP” and compare with two other protocol RDRP and AODV [9]. Saoucene Mahfoudh and Pascale Minet enhanced OLSR to EOLSR by replacing multipoint relays (MPRs) with energy-aware multipoint relays (EMPRs) [10]. In this review paper Neeraj Tantubay, Dinesh Ratan Gautam and Mukesh Kumar Dhariwal present a summary of different energy control techniques and various powers saving methods have been proposed in different research articles [11]. Dr. S. P. Setty and B. Prasad (The author) compares QOS in energy consumption for proactive and reactive routing protocols with the impact of network size [12]. Ved Prakash, Brajesh Kumar and A. K. Srivastava analyze and compare energy efficiency of topology based and location based routing protocols [13]. Feeney L. M. divides the methods which are used in energy efficient awareness routing protocols in ad-hoc networks [14]. In first method when a host transmitting packets, the routing protocol minimized the total energy consumed during transmitting [15], [16], [17]. In second method load balance between hosts to increase the life time of whole network, instead of managing energy consumption for individual packet [18], [19], [20].

Nicolas Chevrollier and Nada Golmie investigate the impact of Bluetooth and wireless standard IEEE 802.15.4 in medical environment. Moreover, they find the importance of both technologies with respect to scalability issues [21].

## 4. SIMULATIONS

### Case I: Methodology Used in Our Simulation-I

The Reference Point Group Mobility model (RPGM) has been used with node speed between 0.5 to 5.0 m/s, simulation time is 900 seconds, transport protocols is UDP and traffic generator source is CBR. The node density and simulation area varies from 20 to 80 nodes and 500mx500m to 2000mx2000m respectively. The initial energy of each node is 1000 joules and two-ray ground is used as propagation model. The other network parameter used in our simulation is described in table 1.

Table 1

| Simulation I Parameters | |
|---|---|
| Parameters | Values |
| MAC Type | IEEE 802.11 |
| Antenna | Omni directional |
| Simulation Time | 900 sec |
| Transmission range | 500 x 500 – 2000 x 2000 |
| Node speed | 0.5m/s to 5.0 m/s |
| Traffic Type | CBR |
| Data payload | 512 bytes/ packet |
| Packet rate | 8 packet/sec |
| Node Pause Time | 0 |
| Mobility Model | RPGM |
| Interface Queue Type | Drop Tail/Priori Queue |
| Interface Queue Length | 50 |
| No. of Nodes | 20 to 80 |

In table 1 some parameters are constant and some are variable. These parameters varying during simulation to test and verify the results (simulation area, node pause time, mobility model and number of node).

### Energy Consumption Model

There are four states of energy consumption of mobile devices which are given in table 2.

Table 2

| Energy Consumption Parameters | |
|---|---|
| ei: | Energy Consumption during Idle mode |
| es: | Energy Consumption during Sleep mode |
| et: | Energy Consumed during Transmitting mode |
| er: | Energy Consumed during Receiving mode |

The fifth parameter Energy consumed during forwarding mode is not used directly because during forwarding mode the host first received packets and send to the next hop. The two parameters er: and et: involved.

### Case II: Methodology Used in Our Simulation-II

The energy model has presented by Santashil Pal Chaudhuri and David B. Johnson in [22], and Dr. S. P. Setty and B. Prasad used this energy model in their paper. The Random waypoint Mobility model has been used with node speed 1 to 10 m/s, simulation time is 300 seconds, transport protocols is UDP and traffic generator source is CBR. The node density varies from 5 to 25 nodes and simulation area 600x600. The initial energy of each node is 1000 joules and two-ray ground is used as propagation model. The other network parameter used in our simulation is described in table 3.

Table 3

| Simulation II Parameters | |
|---|---|
| Parameters | Values |
| MAC Type | IEEE 802.11 |
| Antenna | Omni directional |
| Simulation Time | 300 sec |
| Transmission range | 600 x 600 m |
| Node speed | 1 m/s to 10 m/s |
| Traffic Type | CBR |
| Data payload | 512 bytes/ packet |
| Packet rate | 8 packet/sec |
| Node Pause Time | 0 |
| Mobility Model | Random Waypoint |
| Interface Queue Type | Drop Tail/Priori Queue |
| Interface Queue Length | 50 |
| No. of Nodes | 5, 10, 15, 20, 25 |

We analyze the performance indexes and consumed energy depending on the following operations.

1. Consumed Energy in Rx Mode
2. Consumed Energy in Tx Mode
3. Consumed Energy in Idle Mode
4. Average Remaining Energy
5. Routing Overhead (RO)
6. Packet delivery Ratio / Function (PDR)

7. Average Throughput

**Energy Consumption Model:** The energy consumption model [22], [23], [24] describe total host energy spent in the following modes: (1) TX Mode (2) RX Mode (3) Idle Mode and (4) Overhearing Mode. These modes are describe as under

**1. TX Mode**

When a node send packet to other nodes, it is in TX mode. The energy required during transmit packet is called TX Energy [24], [11] of a node. TX Energy depends on packet size (in bits). TX energy can be described as follows.

$$TX = (\text{Pkt-size} \times 330) / 2 \times 10^6$$

And

$$P_{TX} = TX / T_{TX}$$

Where $P_{TX}$ is transmitting power, TX is transmitting energy and $T_{TX}$ is time take during packet transmit and Pkt-size is the size of packet in bits.

**5. RX Mode**

When a node receives packet from other nodes it is said to be in RX mode. The energy required during receiving packet is called RX energy [25], [26]. The RX energy can be formulated as

$$RX = (\text{Pkt-size} \times 230) / 2 \times 10^6$$

And

$$P_{RX} = RX / T_{RX}$$

Where $P_{RX}$ is receiving power, RX is receiving energy and $T_{RX}$ is time take during receiving a packet and Pkt-size is the size of packet in bits.

**5. Idle/ Listening Mode**

According to idle mode, the node does not send or receive any data packet. But in this mode energy consumed because the node continuously listening the wireless channel and ready to receive packet. When a packet is arrived and the node is converted from idle mode to RX mode. The power consumed in idle mode is as under.

$$P_{Idle} = P_{RX}$$

Where $P_{RX}$ is power consumed in receiving mode and $P_{Idle}$ is power consumed in idle mode.

**5. Drop / Overhearing Mode**

When a packet is receive by a node which is not design for this node it is called overhearing mode. The power consumed in overhearing mode is describe as under.

$$P_O = P_{RX}$$

Where $P_O$ is power consumed in overhearing mode and $P_{RX}$ is power consumed in receiving power.

## 5. RESULTS

### 5.1 Simulation-I

The energy consumption in DSR, DSDV and AODV protocols are evaluated in term of average energy consumed. The node density varies from 20 nodes to 80 nodes.

**1. Energy Consumption in Idle Mode:**

According to the Fig 2, it is proved that energy consumption is maximum in DSR protocol, AODV protocol consumes medium energy and DSDV protocol consumes minimum energy in idle mode.

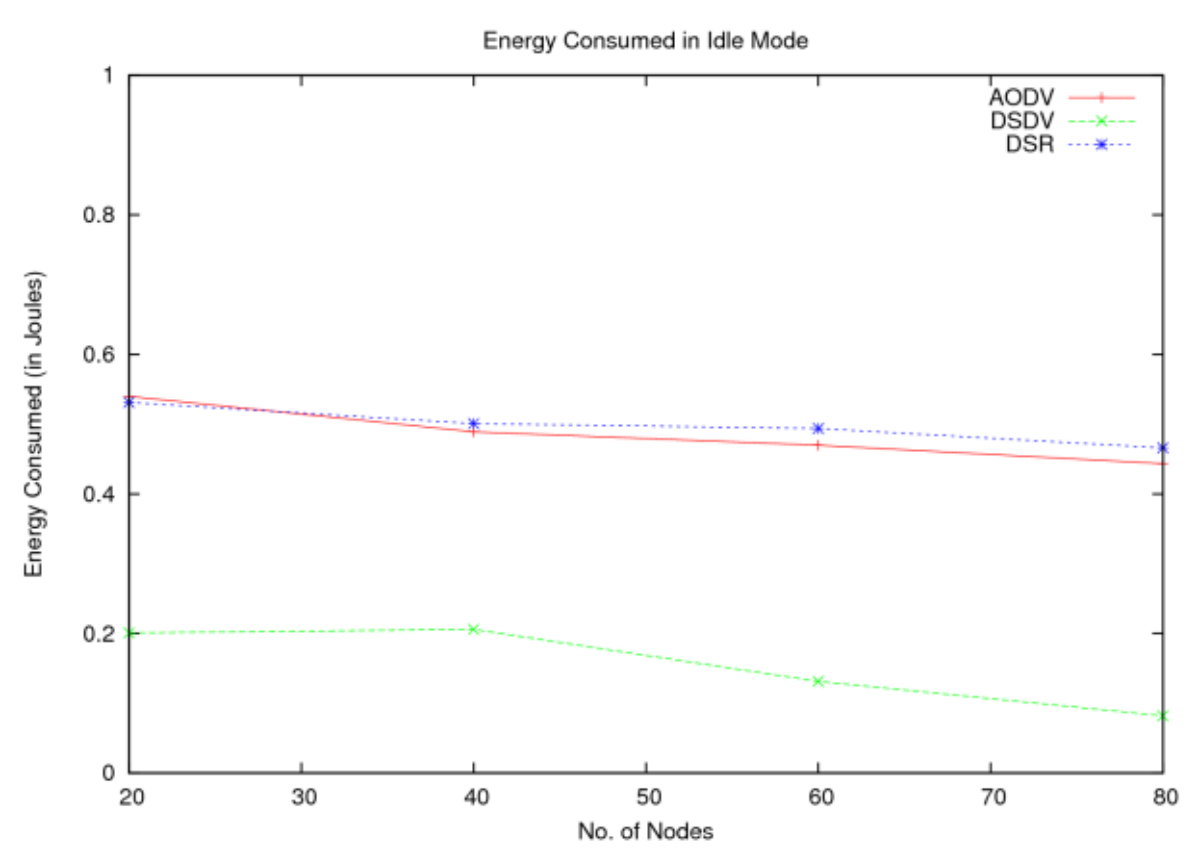

Fig 2: Comparison of average energy consumed in DSDV, DSR and AODV protocols in idle mode.

**2. Energy Consumption in TX Mode:**

According to the Fig 3, it is proved that energy consumption in AODV protocol is maximum DSR protocol consumes medium energy and DSDV protocol consumes minimum energy in TX mode.

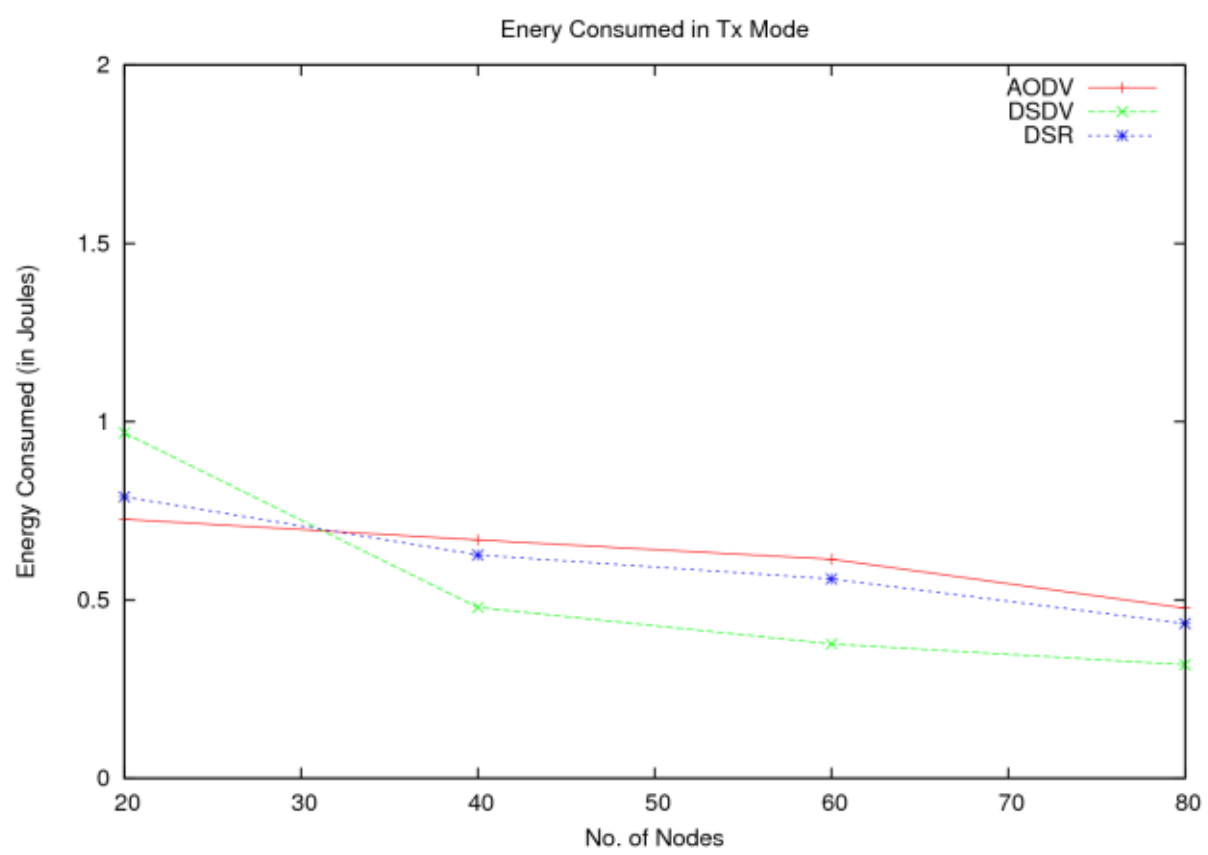

Fig 3: Comparison of average energy consumed in DSDV, DSR and AODV protocols in TX mode.

**7. Energy Consumption in RX Mode:**

According to the Fig 4, it is interesting that energy consumption of DSR in RX mode is less among all the three routing protocols. DSDV protocol consumes minimum energy in idle and TX mode and consumes maximum energy in RX mode. Because DSDV is a proactive protocol so it update table periodically.

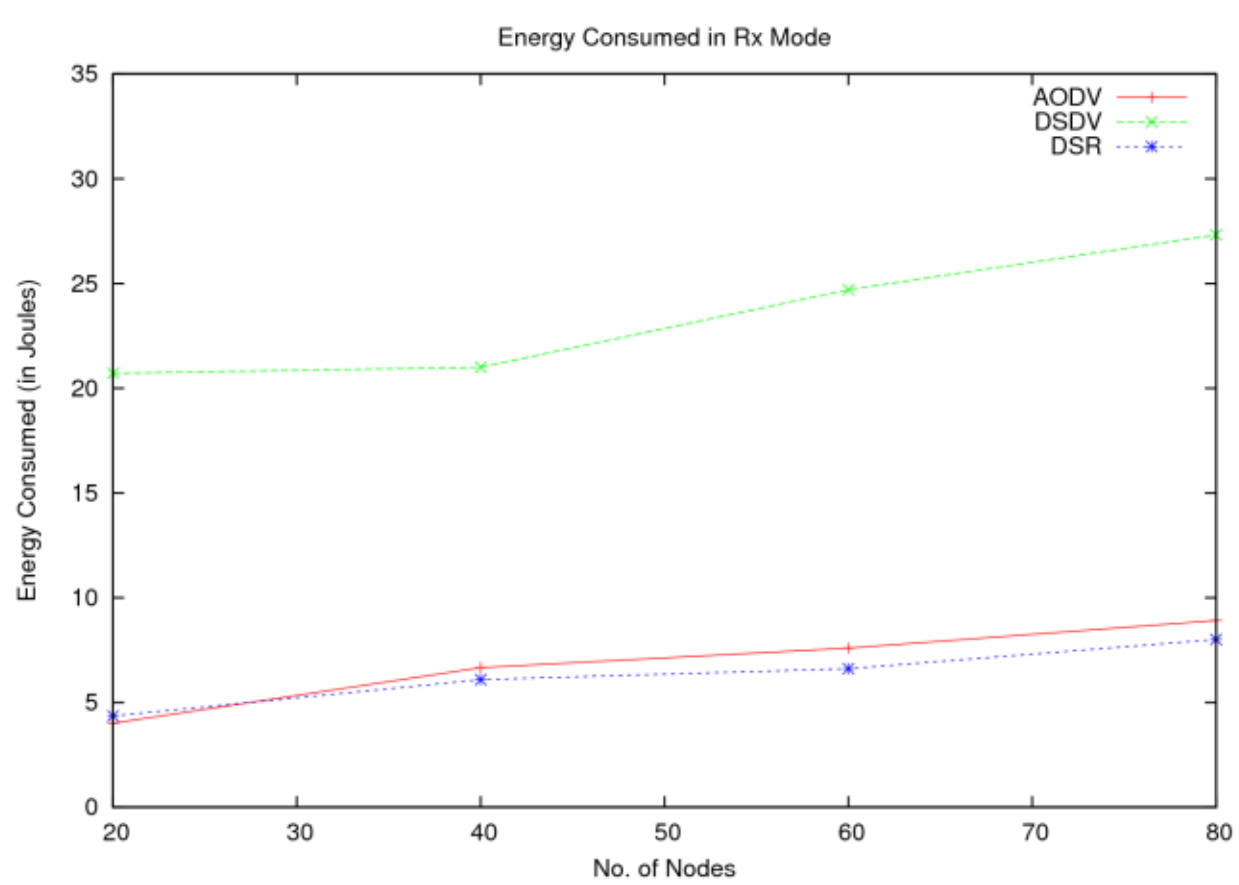

Fig 4: Comparison of average energy consumed in DSDV, DSR and AODV protocols in RX mode.

**5. Average Remaining Energy:**

According to the Fig 5, it is experimental prove that average remaining energy in DSDV protocol is minimum and medium in AODV and maximum in DSR protocol. It means that the performance of DSR is best in this scenario.

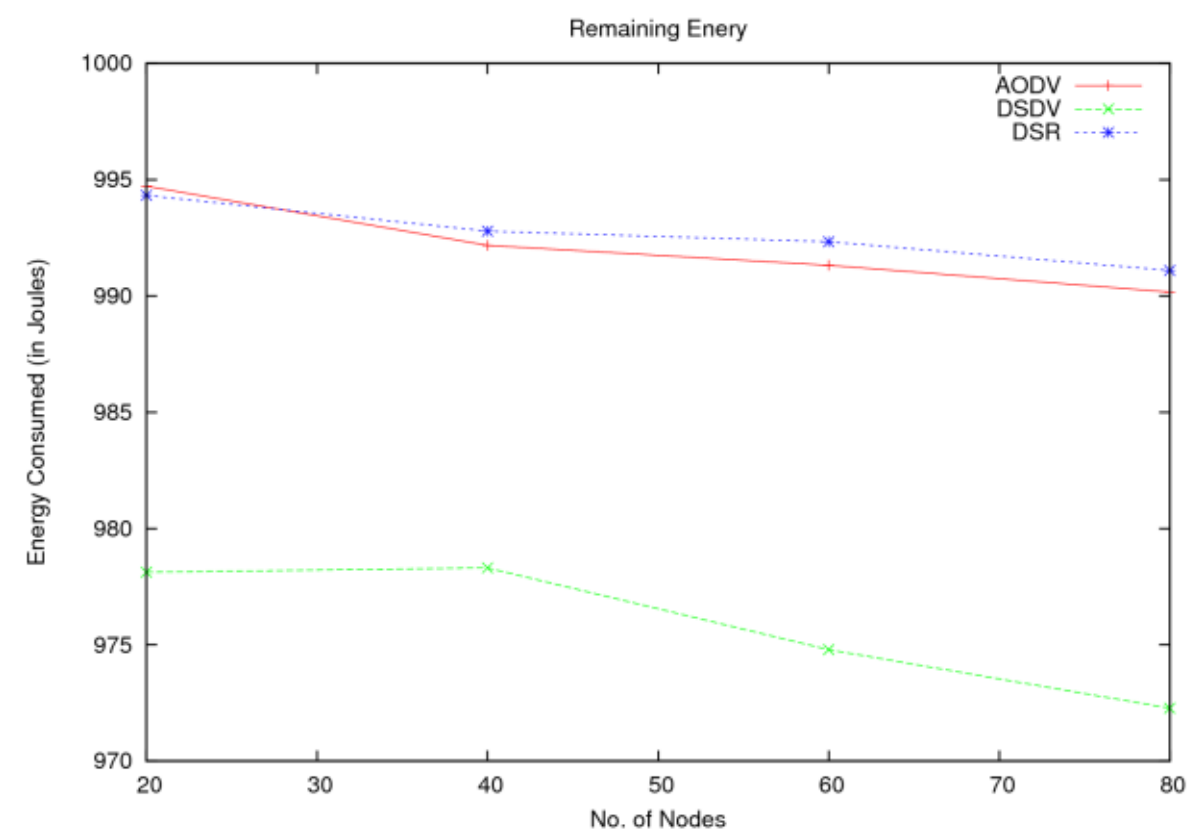

Fig 5: Average remaining energy.

## 5.2 Simulation-II

The energy consumption in DSR, DSDV and AODV protocols are evaluated in term of average energy consumed. The node density varies from 5 nodes to 25 nodes.

**1. Energy Consumption in Idle Mode:**

According to the Fig 6, it is experimental prove that energy consumption in DSDV protocol is maximum, AODV protocol consumes medium energy and DSR protocol consumes minimum energy in idle mode.

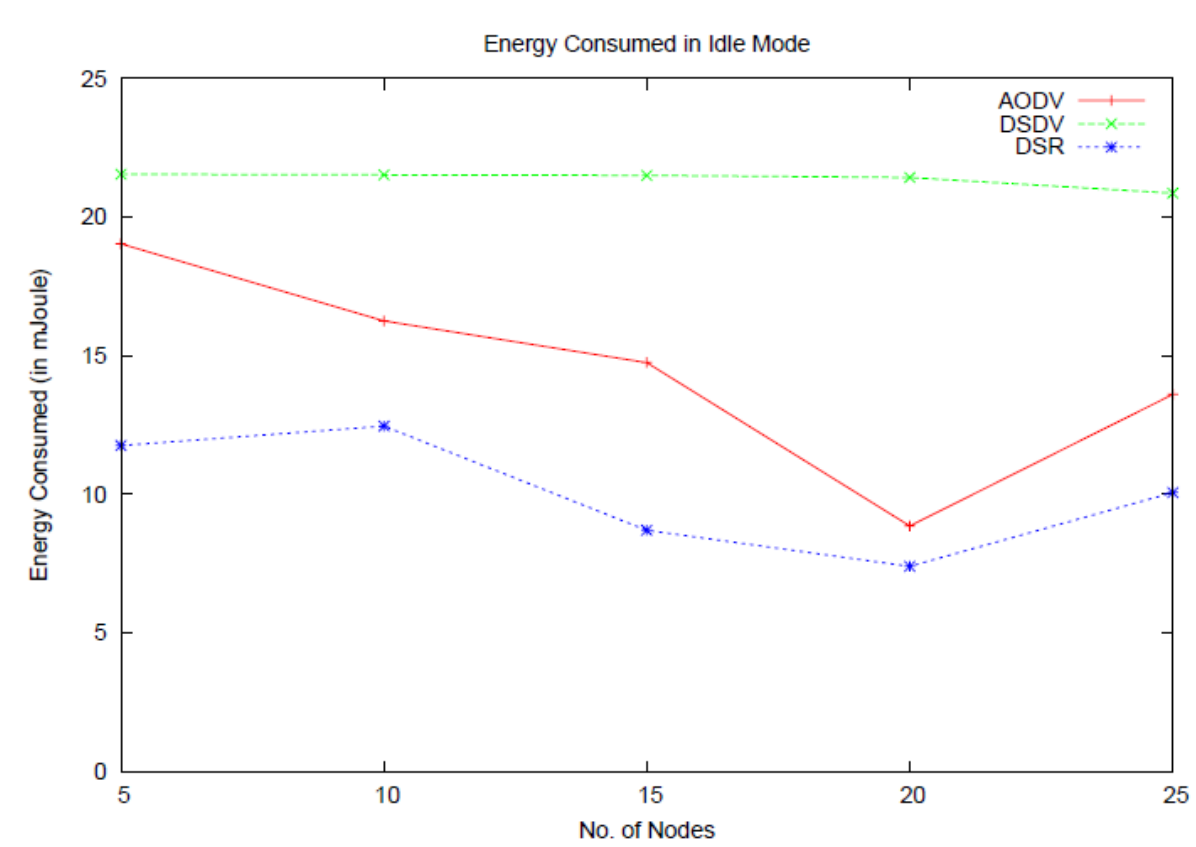

Fig 6: Comparison of average energy consumed in DSDV, DSR and AODV protocols in idle mode.

**2. Energy Consumption in TX Mode:**

According to the Fig 7, it is proved that energy consumption in AODV protocol is maximum DSR protocol consumes medium energy and DSDV protocol consumes minimum energy in TX mode.

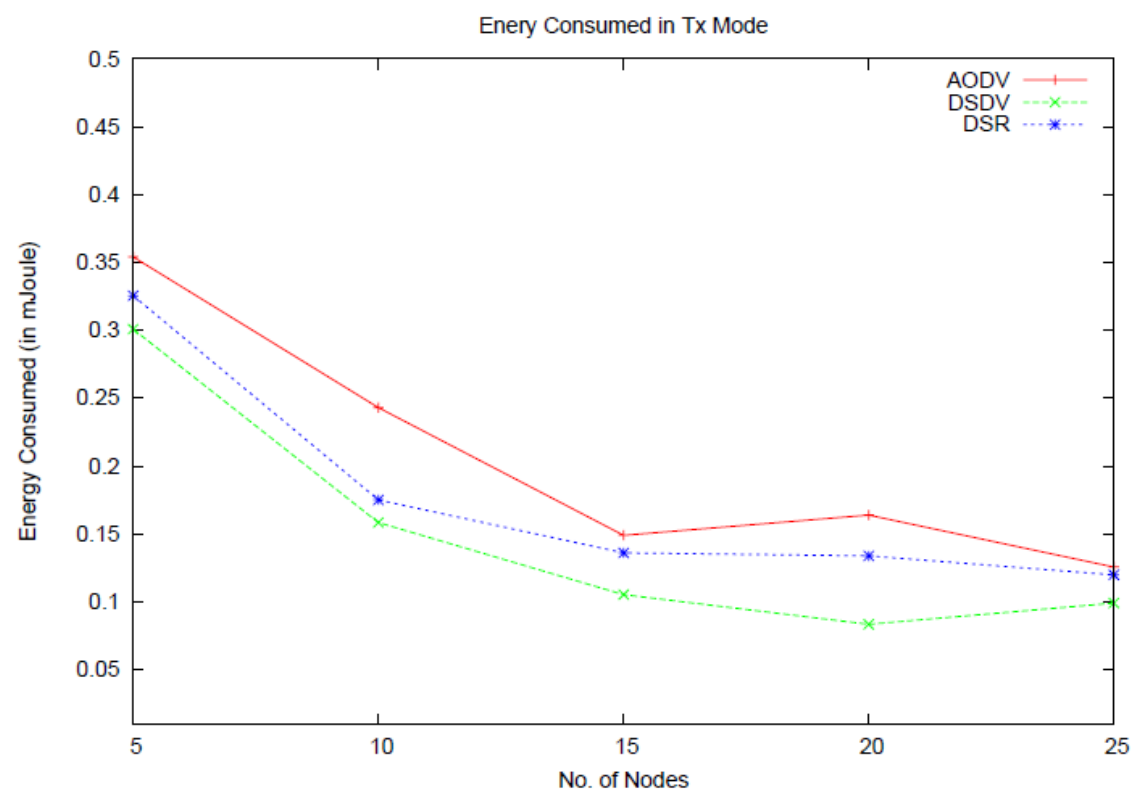

Fig 7: Comparison of average energy consumed in DSDV, DSR and AODV protocols in TX mode.

### 7. Energy Consumption in RX Mode:

The energy consumption of DSR in RX mode is less among all the three routing protocols in Fig 8. DSDV protocol consumes medium energy and AODV consumes maximum energy.

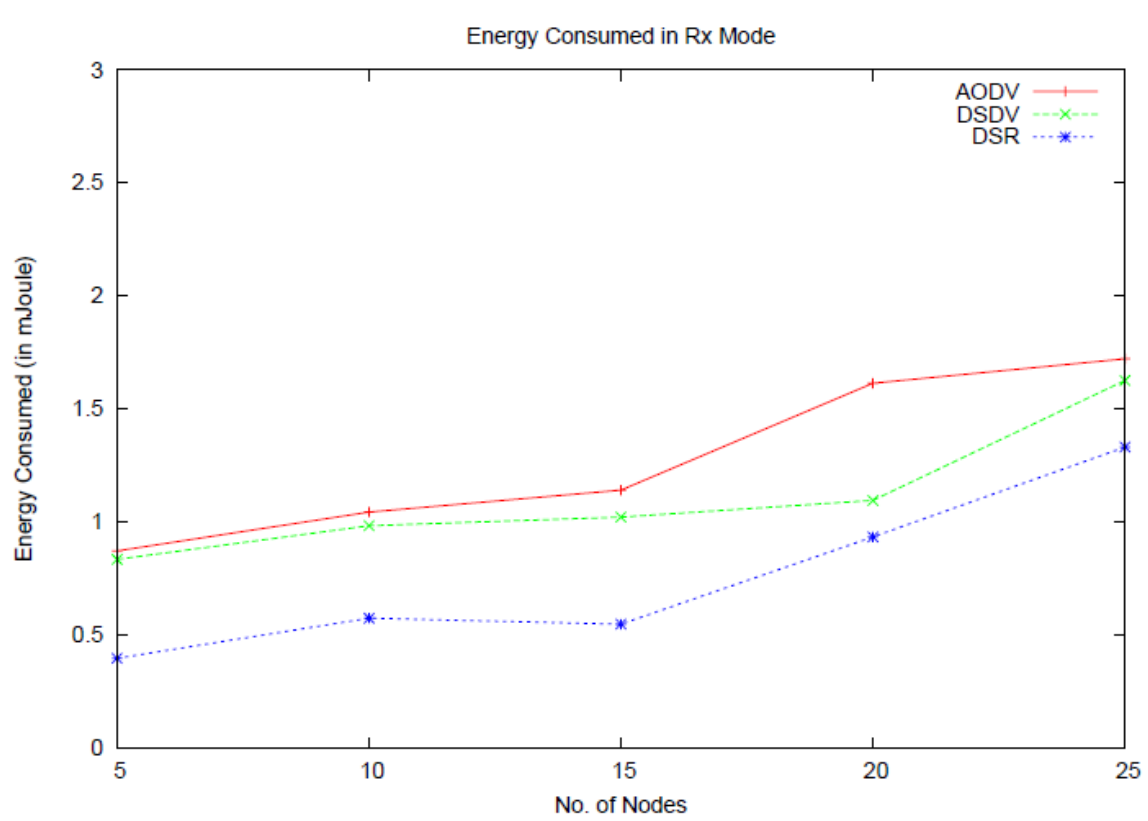

Fig 8: Comparison of average energy consumed in DSDV, DSR and AODV protocols in RX mode.

### 4. Average Remaining Energy:

According to the Fig 9, it is proved that average remaining energy in DSDV protocol is minimum and medium in AODV and maximum in DSR protocol. It means that the performance of DSR is best in this scenario.

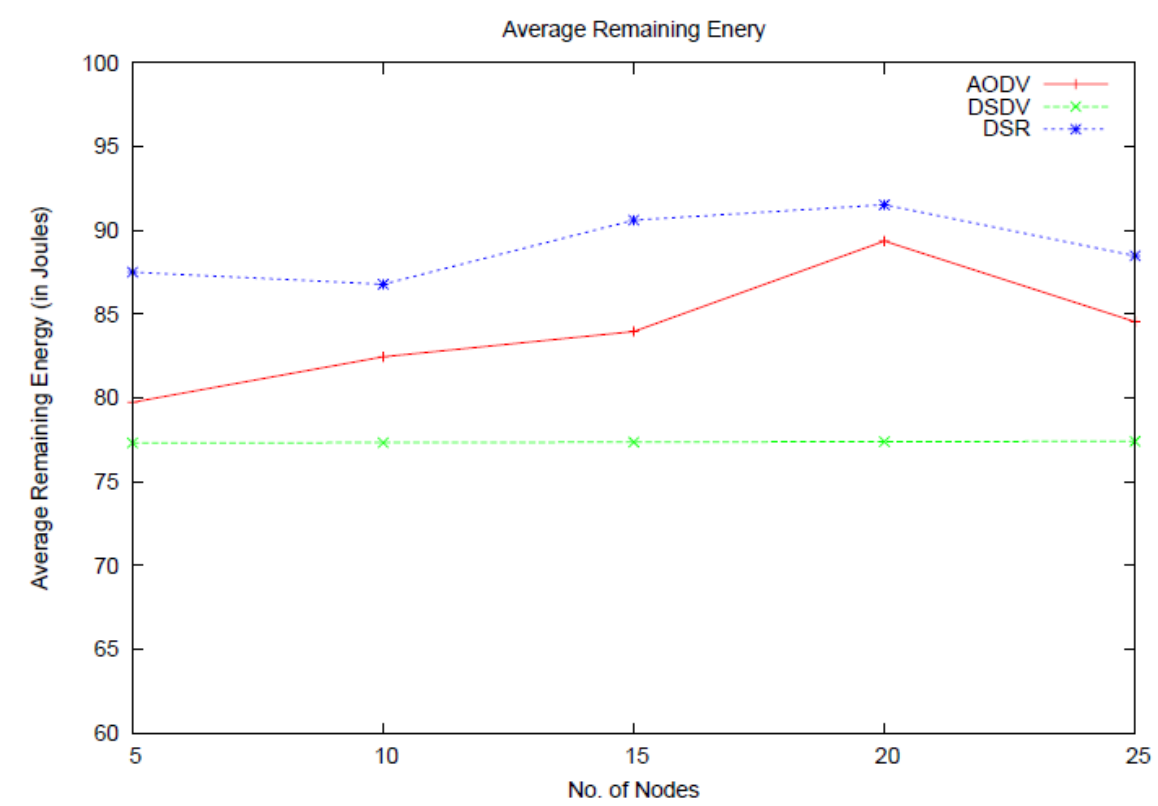

Fig 9: Average remaining energy.

### 5. Routing Overhead (RO):

According to the fig 10, it is analyzed that DSDV has maximum routing overhead AODV has medium but very close to DSR and DSR has minimum RO.

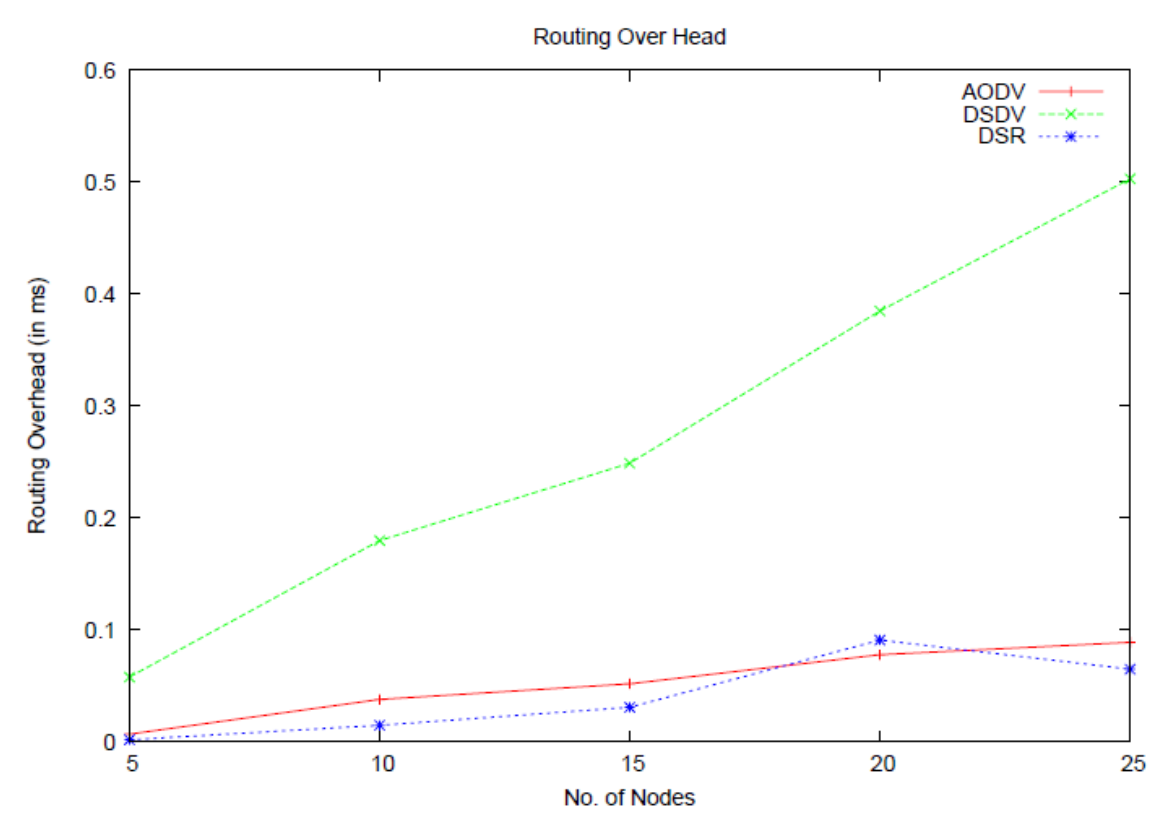

Fig 10: Routing Overhead.

### 6. Packet Delivery Ratio:

According to the fig 11, it is analyzed that DSDV has less packet delivery ratio AODV has medium but very close to DSR and DSR has maximum PDR.

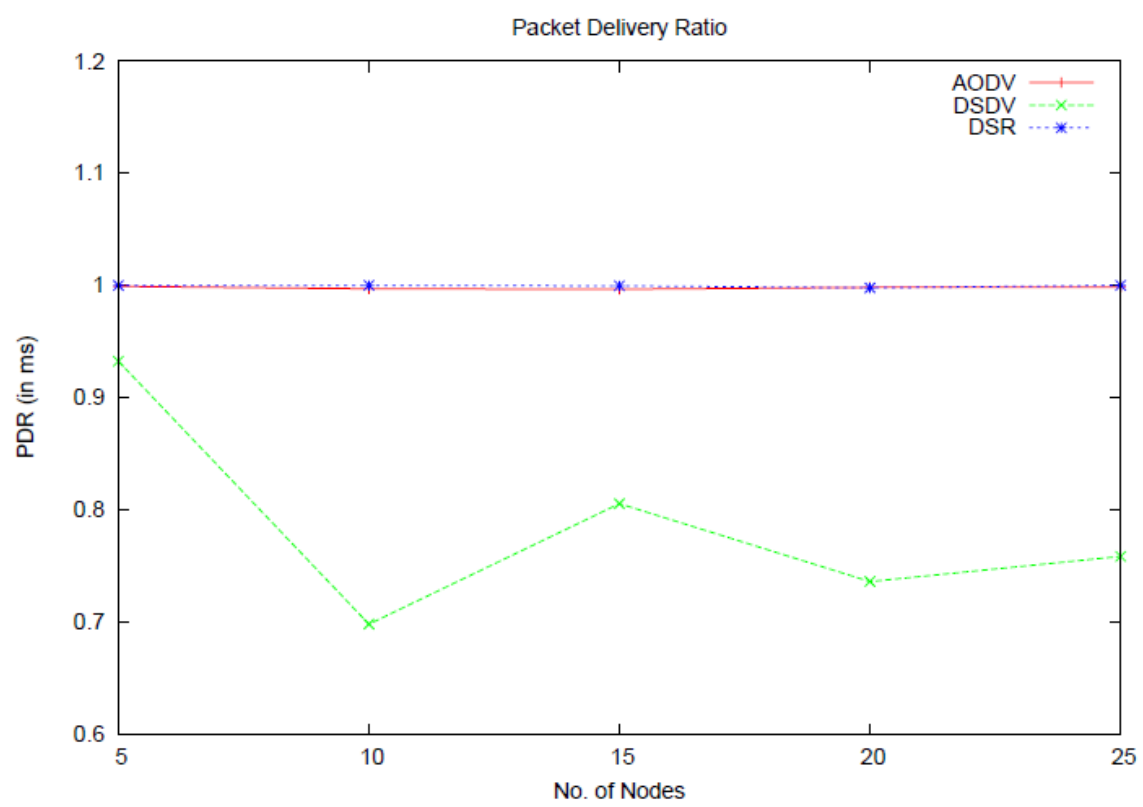

Fig 11: Packet Delivery Ratio.

### 7. Average Throughput:

According to the fig 12, it is analyzed that DSDV has less throughput AODV has medium but close to DSR and DSR has maximum throughput.

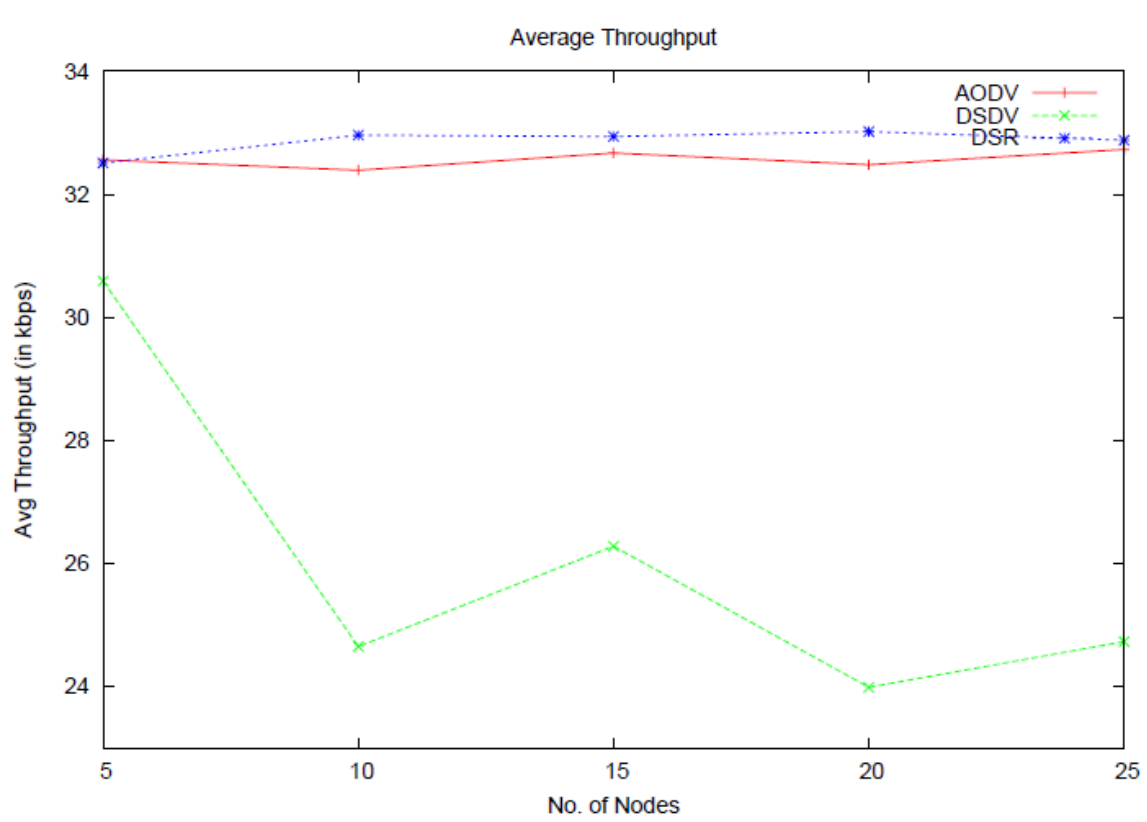

Fig 12: Average Throughput.

## CONCLUSION

In this paper we simulate and compare three routing protocols to investigate the energy consumption in healthcare environment and find in both cases during Idle and TX mode performance of DSR is best when node density is less than and equal to 20 and when node density is greater than 20 the performance of DSDV is best as an energy efficient routing protocol. It is according to average remaining energy graph DSR is maximum remaining energy protocol in both cases On the other hand the performance of DSR is also good on the basis of RO, PDR and throughput.

## Acknowledgments

I greatly acknowledged the kind supervision of Dr. Imran Shafi, who taught me the subject Mathematical Methods in Communication, ANN and encouraged me for this work and to write this paper.

## References

[1] N. A. Pantazis, S. A. Nikolidakis, D. D. Vergados, " Energy-efficient routing protocols in wireless wensor networks for health communication systems", PETRA 09, Corfu, Greece, ACM ISBN 978-1-60558-409-6, 2009.
[2] P. Bergamo, Alessandra, " Distributed power control for energy efficient routing in ad hoc networks ", Wireless Networks 10, pp. 29–42, 2004.
[3] V. Shnayder, B. Chen, K. Lorincz, T. R. F. FulfordJones, M. Welsh, "Sensor networks for medical care", this document is a technical report. It should be cited as: Technical Report TR-08-05, Division of Engineering and Applied Sciences, Harvard University, 2005.
[4] I. Noorzaie, "Survey Paper: Medical applications of wireless networks", last modified: April 12, 2006. Note: This paper is available on-line at http://www.cse.wustl.edu/~jain/cse574-06/ftp/medical_wireless/index.html.
[5] I. F. Akyildiz, I. H. Kasimoglui, "Wireless sensor and actor networks:research challenges", (Elsevier) Journal, pp. 351-367, 2004.
[6] D. B. Johnson, D. A. Maltz, J. Broch. "DSR: The dynamic source routing protocol for multi-hop wireless ad-hoc networks", Ad-hoc Networking, chapter 5, pp. 139-172, 2001.
[7] L. M. Feeney, "An energy consumption model for performance analysis of routing protocols for mobile ad hoc networks", Mobile Networks and Applications 6, pp. 239–249, 2001.
[8] F. Giroire∗ , D. Mazauric∗ , J. Moulierac∗ , B. Onfroy∗ , "Minimizing routing energy consumption: from theoretical to practical result", IEEE / ACM International Conference on Green Computing and Communications, 2010.
[9] L. Layuan, L. Chunlin, Y. Peiyan, "An energy level based routing protocol in ad-hoc networks", proceedings of the IEEE / WIC / ACM International Conference on Intelligent Agent Technology (IAT), 2006.
[10] S. Mahfoudh, P. Minet, "Energy-aware routing in wireless ad hoc and sensor networks", IWCMC, Caen, France, ACM 978-1-4503-0062-9/10/06, 2010.
[11] N. Tantubay, D. R. Gautam, M. K. Dhariwal , "A review of power conservation in wireless mobile ad-hoc network (MANET)", IJCSI International Journal of Computer Science Issues, ISSN (Online), Vol. 8, Issue 4, No 1, 2011.
[12] S. P. Setty, B. Prasad, "Comparative study of energy aware QoS for proactive and reactive routing protocols for mobile ad-hoc networks", International Journal of Computer Applications, Volume 31, No.5, 2011.
[13] V. Prakash, B. Kumar, A. K. Srivastava, "Energy efficiency comparison of some topology-based and location-based mobile ad-hoc routing protocols", ICCCS 11, Copyright © 2011 ACM 978-1-4503-0464-1/11/02, Rourkela, Odisha, India, 2011.
[14] Feeney, "Energy efficient communication in ad hoc wireless networks",

http://citeseerx.ist.psu.edu/viewdoc/summary?doi=10.1.1.14.9379.
[15] Bergamo, Giovanardi, Travasonia, "Distributed power control for energy efficient routing in Ad-hoc a networks", published in journal of Wireless Networks, Volume 10 Issue 1, pp. 29–42, 2004.
[16] Gomez, Campbell, Naghshineh, "Conserving transmission power in wireless ad-hoc networks ", [A]. In Proc of IEEE Conference on Network Protocols (ICNP. 01), 2001.
[17] Y. Wei, L. Jangwon, "DSR-based energy-aware routing protocols in ad hoc networks", in Proc. of the International Conference on Wireless Networks, 2002.
[18] W. Zhao, K. Ramamritham, "Virtual time CSMA protocols for hard real–time communications", IEEE Transactions on Software Engineering, Vol. 13, No. 8, 1987.
[19] W. Zhao, J. Stankovic, K. Ramamritham, "A window protocol for transmission of time constrained messages", IEEE Transactions on Computers, Vol. 39, No. 9, 1990.
[20] Rahman, Gburzynski, "On constructing minimum-energy path-preserving graphs for ad-hoc wireless networks", ICC 2005: IEEE international conference on communications, Vols.1, pp. 3083-3087, 2005.
[21] N. Chevrollier, N. Golmie, "On the use of wireless network technologies in healthcare environments", White Paper (U.S Department of Commerce) published in 2005. (Note: This paper is available on-line at http://w3.antd.nist.gov/pubs/aswn05.pdf)
[22] S. P. Chaudhuri, D. B. Johnson, "Power mode scheduling for ad-hoc networks", IEEE International Conference on Network Protocols, 2002.
[23] J. H. Chang, L. Tassiulas, "Energy conserving routing in wireless ad-hoc networks", Proc. IEEE INFOCOM, Tel Aviv, Israel, pp. 22-31, 2000.
[24] M. Fotino, "Evaluating energy consumption of proactive and reactive routing protocols in a MANET", Proc. 1st Int. Conf. On Wireless Sensor and Actor Networks, pp. 119-130, 2007.
[25] R. Zheng, R. Kravets, "On-demand power management for ad-hoc networks", published in IEEE INFOCOM, 2003.
[26] T. H. Tie, C. E. Tan, S. P. Lau, "Alternate link maximum energy level ad-hoc distance vector scheme for energy efficient ad-hoc networks routing", In Proceedings of International Conference on Computer and Communication Engineering (ICCCE 2010), Kuala Lumpur, Malaysia, 2010.

## AUTHOR'S PROFILES

**Sohail Abid:** (Mobile No: +92-321-5248497)
Sohail Abid Student of MS (TN) at IQRA University Islamabad and working as System Administrator at Foundation University Institute of Engineering and Management Sciences.

**Dr. Imran Shafi:** (Mobile No: +92-334-5323402)
Dr. Imran Shafi is working as Assistant Professor at Abasyn University Islamabad campus, Pakistan.

**Shahid Abid:** (Mobile No: +92-333-5656413)
Shahid Abid having Master in Computer Science and working as Assistant System Administrator at Foundation University Institute of Engineering and Management Sciences.